\documentclass[12pt]{article}
\usepackage{amssymb,amsmath,cite}
\newcommand\omegA{\tilde\omega}
\newcommand\Rr[1][r]{R_{\omega l}({#1})}
\newcommand\dRr{\dot R_{\omega l}({r})}

\newcommand\Ylm{Y_{lm}(\theta,\varphi)}
\newcommand\Zz[1][z]{Z_{\omega l}(#1)}
\newcommand\dZz{\dot Z_{\omega l}(z)}
\newcommand\ddZz{\ddot Z_{\omega l}(z)}
\newcommand\Hz{h_{\omega l}(z)}
\newcommand\dHz{\dot h_{\omega l}(z)}
\newcommand\ddHz{\ddot h_{\omega l}(z)}
\newcommand\Ll{\lambda_{\omega l}}
\newcommand\al[1][+]{1/2#1i(\omegA\!+\!\Ll)}
\newcommand\bl[1][+]{1/2#1i(\omegA\!-\!\Ll)}
\newcommand\cl[1][+]{1\!#1\!2iz_0\omegA}
\newcommand\Qq{Q_{\omega l}(q)}

\newcommand\ddQq{\ddot Q_{\omega l}(q)}
\newcommand\Wz{W_{\omega l}(z)}

\newcommand\ddWz{\ddot W_{\omega l}(z)}

\long\def\@makefntext#1{\parindent 0cm\noindent \hbox to
1em{\hss$^{\@thefnmark}$}#1}

\begin{document}
\begin{titlepage}
\vspace{.5in}
\begin{flushright}
\end{flushright}
\vspace{.5in}
\begin{center}
{\Large\bf
 Hawking radiation of asymptotically non-flat dyonic black holes in Einstein-Maxwell-dilaton gravity }\\
\vspace{.4in}
{Peter~I.~ Slavov\footnote{\it email: pslavov@phys.uni-sofia.bg}, Stoytcho~ S.~Yazadjiev\footnote{\it email: yazad@phys.uni-sofia.bg}\\
       {\footnotesize\it  \it Department of Theoretical Physics, Faculty of Physics,}
       {\footnotesize \it Sofia University, Sofia, 1164, Bulgaria }\\
            }
\end{center}

\vspace{.5in}

\begin{abstract}
In the present paper we investigate the Hawking radiation of
asymptotically non-flat dyonic black holes in 4D  Einstein-Maxwell-dilaton gravity in semi-classical approximation. We show that the problem allows an exact
analytical treatment and we compute exactly the semi-classical radiation spectrum of both non-extremal and extremal black holes under consideration.
In the high frequency regime  we find that the Hawking temperature does not agree with the surface gravity when the magnetic charge is nonzero.
Even more surprisingly  the Hawking temperature is independent  of the black hole intrinsic characteristics, as the mass and magnetic charge, and
depends only on the linear dilaton background parameter.
\end{abstract}

PACS:
\end{titlepage}
\addtocounter{footnote}{-1}

\section{Introduction}

The Hawking radiation of black holes is an emblematic effect in the
quantum field theory in curved spacetime \cite{Hawking},\cite{BD}.
This effect, lying on the wedge of classical and quantum gravity,
reveals  the deep connection between the black hole physics and
thermodynamics. In the semi-classical approximation, the spectrum of
the Hawking radiation can be obtained by computing the Bogoliubov
coefficients in two different vacua and matching them appropriately
\cite{BD}. Another procedure which, in asymptotically  flat
spacetimes,  gives the same result as the Bogoliubov coefficients
method, is to compute the absorption and the transmission (or
reflection) coefficients of waves defined at the asymptotic regions
\cite{Unruh}.

Ever since its discovery the Hawking radiation continued to be a hot area
of research. The importance of the Hawking effect for the fundamental physics
stimulates its investigation for various  black hole solutions with different
structures and asymptotics.  Unfortunately, the wave equations in black hole spacetimes can not be solved
analytically and this makes the full study of the Hawking radiation hard. Only in special cases we are able
to solve the wave equations exactly. The cases when the wave equation is exactly solvable are important because
they enable us to study the Hawking radiation in detail and, for example,  allow us to compute the radiation spectrum exactly.
In the present paper we consider one such case when the exact analytical treatment of the Hawking radiation is possible.
More precisely we study the Hawking radiation of a class of asymptotically non-flat dyonic black holes in the 4D Einstein-Maxwell-dilaton gravity.

 The Einstein-Maxwell-dilaton gravity in 4-dimensional spacetime is
described by the following action
\begin{eqnarray}\label{EMDA}
S= \frac{1}{16\pi G} \int d^4x \sqrt{-g}\left[{\cal R} -
2g^{\mu\nu}\nabla_{\mu}\varphi \nabla_{\nu}\varphi -
e^{-2\alpha\varphi}F_{\mu\nu}F^{\mu\nu}\right] ,
\end{eqnarray}
where ${\cal R}$ is the scalar curvature with respect to the spacetime
metric $g_{\mu\nu}$, $F_{\mu\nu}$ is the electromagnetic field, and
$\varphi$ is the scalar dilaton field with a coupling constant
$\alpha$.  In the present paper we are interested in the black hole
solutions of (\ref{EMDA}). More precisely we consider the following
dyonic black hole  solution for $\alpha=1$ found in \cite{Y}:

\begin{eqnarray}\label{dS^2}
&& ds^2= - \frac{(r-r_{-})(r-r_{+})}{r_{0}r}dt^2 +
\frac{{r_{0}r}}{(r-r_{-})(r-r_{+})}dr^2 + r_{0}r\left(d\theta^2 +
\sin^2\theta d\phi^2\right), \\
&&\Phi_{e}=\frac{1}{\sqrt{2}}\frac{r}{r_{0}},\\
&&\Phi_{m}=\sqrt{\frac{r_{-}}{2r_{+}}} \frac{r_{+}}{r},\\
&&e^{2\varphi}=\frac{r}{r_{0}} ,
\end{eqnarray}
where $r_{-}$, $r_{+}$ and $r_{0}$ are constants. Here $\Phi_{e}$
and $\Phi_{m}$ are the electric and magnetic potentials and the
Maxwell 2-form is given by

\begin{eqnarray}
F= d\Phi_e \wedge dt   + e^{2\varphi}\star \left(d\Phi_{m}\wedge dt
\right)
\end{eqnarray}
with $\star$ being the Hodge dual.

The solution describes an asymptotically non-flat dyonic black hole
with  inner and outer horizons at $r=r_{-}$ and $r_{+}$,
respectively. The electric and magnetic charges are given by

\begin{eqnarray}
Q= \frac{r_{0}}{\sqrt{2}}, \; \; \; P=\sqrt{\frac{r_{+}r_{-}}{2}}.
\end{eqnarray}

In the limit $P\to 0$ ($r_{-}\to 0$) we recover the pure electrical linear dilaton black hole solution \cite{CHM} (see also \cite{CGL}).
It should be noted that the parameter $r_{0}$ (or equivalently the electric charge $Q$)
is associated not with a specific black hole, but rather with a given linear dilaton background.  The linear dilaton background solution
is obtained by setting $r_{+}=r_{-}=0$ in our solution.

In order to find the physical mass of the black hole solution under consideration we use
the quasilocal formalism \cite{BY}. Since our spacetime is not asymptotically flat, a suitable
substraction procedure is needed to obtain a finite mass. In our case the linear dilaton background
is the most  natural and unique choice for the substraction background. The explicit calculations give the following
result for the mass \cite{Y}

\begin{eqnarray}
M= \frac{1}{4}\left(r_{+} + r_{-}\right).
\end{eqnarray}

The surface gravity is given by

\begin{eqnarray}\label{surface_gravity}
\kappa = \frac{r_{+} - r_{-}}{2 r_{0} r_{+}}  .
\end{eqnarray}

In the case with $r_{+}=r_{-}$ we obtain an extremal black hole solution with zero surface gravity.

Following \cite{CGLO} one  can formally derive the first law for the
black holes under considerations, namely

\begin{eqnarray}\label{FLaw}
dM= \frac{\kappa}{2\pi} d\left(\frac{A_{\cal H}}{4}\right) +
\Phi_{m}^{{\cal H}}dP
\end{eqnarray}
where $A_{\cal H}=4\pi r_{0}r_{+}$ is the horizon area and
$\Phi_{m}^{{\cal H}}=\sqrt{\frac{r_{-}}{2r_{+}}}$ is the magnetic
potential evaluated on the horizon $r=r_{+}$. Here the parameter
$r_{0}$ (respectively $Q$) is kept fixed since it is associated with
the background. General comments about the black hole thermodynamics
in asymptotically non-flat spacetimes will be given in the last
section.

In the next section we study the Hawking radiation of  the
asymptotically non-flat black hole solutions  presented above in
semi-classical approximation in both the non-extremal and extremal
case. The Hawking radiation of pure electrical linear dilaton black
holes (i.e. corresponding to  $P=0$) was studied in semi-classical
approximation in \cite{CFM}.

\section{Hawking  radiation in semi-classical
approximation}

\subsection{Hawking radiation of non-extremal black holes }

In order to study the Hawking radiation of our black holes we consider a test scalar field $\psi$ satisfying the wave equation
\begin{eqnarray}\label{WE}
\Box \psi=0 ,
\end{eqnarray}
where \ $\Box$ \ is the curved spacetime D'alambert operator.

For the static spherically-symmetric metric (\ref{dS^2}), the D'alambert operator takes the following explicit form:
$$\Box=\frac1{\sqrt{\!-g}}\partial_\mu\!\left(\sqrt{\!-g}\,g^{\mu\nu}\partial_\nu\;\;\right)
=-\frac{r_0\,r}{(r\!-\!r_-)(r\!-\!r_+)}\partial^2_{tt}+\frac1{r_0r}\partial_r[(r\!-\!r_-)(r\!-\!r_+)\partial_r\;\;]+
\frac1{r_0r}\Delta_{(\theta,\varphi)}\, ,
$$
where $\Delta_{(\theta,\varphi)}$ is the Laplace operator on the unit sphere $S^2_{r=1}$.
So the scalar wave equation \ $\Box\psi=0$ \ multiplied by \ $r_0r$ \ becomes:
\begin{equation}\label{wave}
-\frac{r_0^2r^2}{(r\!-\!r_-)(r\!-\!r_+)}\partial^2_{tt}\psi+\partial_r\left[(r\!-\!r_-)(r\!-\!r_+)\partial_r\right]\psi+\Delta_{(\theta,\varphi)}\psi=0\,.
\end{equation}

Consider now a harmonic eigenmode as a partial solution of (\ref{wave}) in separate variables:
\begin{equation}
\psi_{\omega lm}(t,r,\theta,\varphi):=\Rr\Ylm e^{-i\omega t}\,
\end{equation}
where $Y_{lm}(\theta,\varphi)$ are the spherical harmonics. Then the Fourier coefficients $\Rr$ satisfy the equation:
\begin{equation}\label{R-wave}
\frac{d}{dr}\left[(r\!-\!r_-)(r\!-\!r_+)\dRr\right]+\left[\frac{\omegA^2r^2}{(r\!-\!r_-)(r\!-\!r_+)}-l(l\!+\!1)\right]\Rr=0\,,
\end{equation}
where $\omegA:=r_0\omega$ is a dimensionless frequency.

After the substitutions:
\begin{equation}\label{r-subst}
z:=\frac{r_+\!-r}{r_+\!-\!r_-}\;,\quad z_0:=\frac{r_+}{r_+\!-\!r_-}>1
\end{equation}
the equation for $\Zz:=\Rr$ becomes:
\begin{equation}\label{Z-equ}
z(1\!-\!z)\ddZz+(1\!-\!2z)\dZz+\left[\frac{\omegA^2(z\!-\!z_0)^2}{z(1-z)}+l(l\!+\!1)\right]\Zz=0\,.
\end{equation}
An appropriate substitution like $\Zz:=z^p(z\!-\!1)^q \Hz$ will help us to obtain a more familiar linear differential equation:
$$\begin{array}{l}
   z(1\!-\!z)\ddHz+[1\!+\!2p-2(p\!+\!q\!+\!1)z]\,\dHz-\\
   \vspace{-2ex}\\
  \displaystyle\quad -\left[(p\!+\!q\!+\!1)(p\!+\!q)+\omegA^2\!+l(l\!+\!1)-\frac{p^2\!\!+\!z_0^2\omegA^2}z-
   \frac{q^2\!\!+\!(z_0\!-\!1)^2\omegA^2}{1\!-\!z}\right]\!\Hz=0\,
  \end{array}
$$
where \ $p$ \ and \ $q$ \ are determined by the conditions which eliminate the simple rational fraction $1/z$ and $1/(1\!-\!z)$
in the coefficient in front of $\Hz$.
There are four combinations for $p$ and $q$, that can do it, but the most convenient one is to take \ $p:=iz_0\omegA$ \ and \ $q:=-i(z_0\!-\!1)\omegA$:
\begin{equation}\label{hyperg1}
    z(1\!-\!z)\ddHz+[\cl-2(1\!+\!i\omegA)z]\,\dHz-[i\omegA-l(l\!+\!1)]\,\Hz=0\,.
\end{equation}
This is just the hypergeometric equation and after identifying (\ref{hyperg1}) with the canonical form:
\begin{equation}\label{hypergeom}
z(1\!-\!z)\ddHz+\left[c-(1\!+\!a\!+\!b)z\right]\dHz-ab\,\Hz=0\,,
\end{equation}
one can easily obtain the canonical parameters:
\begin{equation}\label{hyper-param}
      \begin{array}{|l}
      a=\al\\
      b=\bl\\
      c=\cl
    \end{array}\;,\quad\textrm{where }\lambda_{\omega l}:=\sqrt{\omegA^2-(l\!+\!1/2)^2}\,.
\end{equation}

In our dimensionless variable (\ref{r-subst}) the singular point \ $z=1$ \ corresponds to the inner horizon $r=r_-$ while the singular point \ $z=0$ \
corresponds to the outer horizon $r=r_+$. We are interested in the general solution around the outer horizon $r=r_+$ that
can be continued to the spatial infinity $z\to-\infty$ ($r\to+\infty$):
\begin{equation}\label{hyperg_gen_sol}
\Hz=C_1\,F\left(a,b,c;z\right)+C_2\,z^{1-c}F\left(1\!+\!a\!-\!c,1\!+\!b\!-\!c,2\!-\!c;z\right) .
\end{equation}
Consequently for
\begin{equation}\label{Zh}
\Zz:=z^{iz_0\omegA}(z\!-\!1)^{-i(z_0\!-\!1)\omegA}\Hz\,,
\end{equation}
we obtain the expression:
$$\begin{array}{rl}
 \displaystyle\Zz=&\displaystyle(1-z)^{-i(z_0\!-\!1)\omegA}\left[C_1(-z)^{iz_0\omegA}
 F\left(\al,\bl,\cl;z\right)+\right.\\
\\
 \displaystyle+&\left.C_2(-z)^{-iz_0\omegA}
F\left(1/2-i\left[(2z_0\!-\!1)\omegA-\!\Ll\right],1/2-i\left[(2z_0\!-\!1)\omegA+\!\Ll\right],\cl[-];z\right)\right].
  \end{array}
$$
using (\ref{hyper-param}) and (\ref{hyperg_gen_sol}).

 For the asymptotic  of (\ref{hyperg_gen_sol}) near the outer horizon (i.e. $z\to 0_{-}$) the zero order expansion of $F\left(a,b,c;z\right)$
 is enough ($F(a,b,c,0)=1$). Also \ $\lim_{z\to0_{-}}(1\!-\!z)^{-i(z_0\!-\!1)\omegA}=1$.
 Then we have
\begin{equation}\label{Z-asympt}
\Zz[z\!\to\!0_{+}]\simeq
C_1(-z)^{iz_0\omegA}+C_2(-z)^{-iz_0\omegA}\,.
\end{equation}
For physical interpretation it is appropriate to define a new real spatial variable $x$, by the relation:
\begin{equation}\label{rx1}
-z=\frac{r-r_+}{r_+-r_-}:=\exp{\frac{x}{z_0r_0}}\,
\end{equation}
with $r\to r_+ \Rightarrow x\to-\infty$ and  $r\to+\infty\Rightarrow x\to+\infty$.
Now taking into account that $x=x(r)$ is an uniformly growing function and replacing (\ref{rx1}) in (\ref{Z-asympt}),
the asymptotic solution (\ref{Z-asympt}) multiplied by $e^{-i\omega t}$ can be considered as a superposition of an out-going and an in-going wave

$$\Rr[r\to r_+]e^{-i\omega t}\simeq A_{out} e^{i\omega(x-t)}+A_{in} e^{-i\omega(x+t)}\;,\;\textrm{ where }
\begin{array}{|r}
A_{out}:=C_1\\
A_{in}:=C_2 .
\end{array}
$$
In the case $l=0$, $Y_{00}(\theta,\varphi)=1$, and the upper expression is
just the solution for the eigenmode \ $\psi_{\omega00}(t,r,\theta,\varphi)$ \ for \ $r=r_+$.

At the spatial infinity ($r\to+\infty$\,, \ $1/z\to0$), the asymptotic solution can be written by using the known
relation between hypergeometric functions $F(a,b,c;z)$ and $F(a',b',c';1/z)$. Taking only the leading order expansion of $F(a',b',c';1/z)$
with respect to $1/z$, we have the following expressions:
\begin{equation}\label{transf1}
\begin{array}{l}\displaystyle
 F(b,c;z\to-\infty)\simeq\frac{\Gamma(c)\Gamma(b\!-\!a)}{\Gamma(b)\Gamma(c\!-\!a)}(-z)^{-a}+\frac{\Gamma(c)\Gamma(a\!-\!b)}{\Gamma(a)\Gamma(c\!-\!b)}(-z)^{-b},\\
 \\
 \displaystyle(-z)^{1-c}F\left(1\!+\!a\!-\!c,1\!+\!b\!-\!c,2\!-\!c;z\to-\infty\right)\simeq\\
 \vspace{-2ex}\\
\quad\displaystyle=\frac{\Gamma(2\!-\!c)\Gamma(b\!-\!a)}{\Gamma(1\!+\!b\!-\!c)\Gamma(1\!-\!a)}(-z)^{-a}+
\frac{\Gamma(2\!-\!c)\Gamma(a\!-\!b)}{\Gamma(1\!+\!a\!-\!c)\Gamma(1\!-\!b)}(-z)^{-b} .
\end{array}
\end{equation}

Applying the transformations (\ref{transf1}) to (\ref{hyperg_gen_sol}), replacing $\Hz$ in (\ref{Zh}) and finally taking into account that
$\displaystyle\lim_{z\to-\infty}\left[-z/(1\!-\!z)\right]^{i(z_0-1)\omegA}=1\,,$
the  general asymptotic solution becomes:
\begin{equation}\label{W-inf}
\Zz[z\to
-\infty]\simeq\frac{1}{\sqrt{-z}}\left[B_{out}(-z)^{i\Ll}+B_{in}(-z)^{-i\Ll}\right]\,,
\end{equation}
where
$$B_{out}:=C_1\frac{\Gamma(c)\Gamma(a\!-\!b)}{\Gamma(a)\Gamma(c\!-\!b)}+
C_2\frac{\Gamma(2\!-\!c)\Gamma(a\!-\!b)}{\Gamma(1\!+\!a\!-\!c)\Gamma(1\!-\!b)}\,,\;
B_{in}:=C_1\frac{\Gamma(c)\Gamma(b\!-\!a)}{\Gamma(b)\Gamma(c\!-\!a)}+
C_2\frac{\Gamma(2\!-\!c)\Gamma(b\!-\!a)}{\Gamma(1\!+\!b\!-\!c)\Gamma(1\!-\!a)}\,.
$$

Following (\ref{rx1}) and applying the substitution
\begin{equation}
\;k_{\omega l}:=\frac{\Ll}{z_0r_0}\,,
\end{equation}
when $\Ll\in\mathbb R$, the radial function in $x$-variable can be considered again like a superposition of 1D modes with wave vectors
\ $\pm k_{\omega l}$ \ ($(\!-z)^{\pm i\Ll}=e^{\pm ik_{\omega l}x}$):

\begin{equation}
\Rr[r\to
+\infty]\simeq \sqrt{\frac{r_{+} - r_{-}}{r}} \left[B_{out}\,e^{ik_{\omega
l}x}+B_{in}\,e^{-ik_{\omega l}x}\right]\;,
\end{equation}

\begin{equation}\label{B1}
\begin{array}{rl}\displaystyle
 B_{out}=\Gamma(2i\Ll)\!\!&\!\!\displaystyle\left[\frac{\Gamma(1+2iz_0\omegA)\,A_{out}}{\Gamma(\al)\,\Gamma(1/2+i[(2z_0\!-\!1)\omegA+\Ll])}\right.\,+\quad\\
 \vspace{-2ex}\\
 &\displaystyle\quad\quad+\left.\frac{\Gamma(1-2iz_0\omegA)\,A_{in}}{\Gamma(\bl[-])\,\Gamma(1/2\!-\!i\left[(2z_0\!-\!1)\omegA\!-\!\Ll\right])}\right],\\
 \\
 \displaystyle
 B_{in}=\Gamma(-2i\Ll)\!\!&\!\!\displaystyle\left[\frac{\Gamma(1+2iz_0\omegA)\,A_{out}}{\Gamma(\bl)\,\Gamma(1/2+i[(2z_0\!-\!1)\omegA\!-\!\Ll])}\right.\,+\quad\\
 \vspace{-2ex}\\
 &\displaystyle\quad\quad+\left.\frac{\Gamma(1-2iz_0\omegA)\,A_{in}}{\Gamma(\al[-])\,\Gamma(1/2\!-\!i\left[(2z_0\!-\!1)\omegA\!+\!\Ll\right])}\right].
\end{array}
\end{equation}

Black hole radiation is considered as a specific boundary condition when only an out-going mode at the spatial infinity exists, $B_{in}=0$.
This  condition determines the ratio of the coefficients $A_{in}/A_{out}$ or the reflection coefficient $R$:
$$R=\left.\frac{|A_{in}|^2}{|A_{out}|^2}\right|_{B_{in}=0}=
\frac{|\Gamma(\cl)|^2|\Gamma(\al[-])|^2|\Gamma(1/2-i[(2z_0\!-\!1)\omegA+\Ll])|^2}%
{|\Gamma(\cl[-])|^2|\Gamma(\bl)|^2|\Gamma(1/2+i[(2z_0\!-\!1)\omegA-\Ll])|^2}.$$

Complex conjugation and the Euler's reflection formula for the Gamma function give us the final result for the reflection coefficient on the outer horizon:
\begin{equation}\label{R}
R=
\frac{\cosh(\pi(\omegA\!-\!\Ll))\cosh(\pi((2z_0\!-\!1)\omegA\!-\!\Ll))}{\cosh(\pi(\omegA\!+\!\Ll))\cosh(\pi((2z_0\!-\!1)\omegA\!+\!\Ll))} .
\end{equation}

In the special case $r_-=0\Rightarrow z_0=1$ we recover the result of  \cite{CFM}
    \begin{equation}\label{Z0=0}
    R=\frac{\cosh^2(\pi(\omegA\!-\!\Ll))}{\cosh^2(\pi(\omegA\!+\!\Ll))}\,.
    \end{equation}

For high frequencies $\omegA\gg l\!+\!1/2$ ($\Rightarrow\Ll\approx\omegA$)  and $\omegA\gg\frac1{z_0\!-\!1}$ we obtain

\begin{eqnarray}
N:&=&\frac{R}{1-R}=\left[\frac{\cosh(2\pi\omegA)\cosh(2\pi z_0\omegA)}{\cosh(2\pi(z_0\!-\!1)\omegA)}-1\right]^{-1}
\!\!=2\!\left[\frac{\cosh(2\pi(z_0\!+\!1)\omegA)}{\cosh(2\pi(z_0\!-\!1)\omegA)}-1\right]^{-1}=
\nonumber \\
&&2\!\left[e^{4\pi\omegA}\frac{1+\exp(-4\pi(z_0\!+\!1)\omegA)}{1+\exp(-4\pi(z_0\!-\!1)\omegA)}-1\right]^{-1} \approx e^{-4\pi\omegA}.
\end{eqnarray}

We identify the Hawking temperature from $N\approx e^{-4\pi\omegA} = e^{-\frac{\omega}{\,T_H}}$ which gives

\begin{eqnarray}
T_{H}= \frac{1}{4\pi r_{0}}.
\end{eqnarray}

As one can see  the black hole temperature derived in semiclassical
approximation  does not agree with the surface  gravity, i.e.
$T_{H}\ne \frac{\kappa}{2\pi}$ where the surface gravity $\kappa$ is
given by (\ref{surface_gravity}). Only for $P=0$  ($r_{-}=0$) we
have $T_{H}= \frac{\kappa}{2\pi}$.

\subsection{Hawking radiation of extremal black holes}

The extremal case can formally be considered as a limit of the non-extremal one, namely in the limit $r_{+}\to r_{-}$, i.e. $z_{0}\to \infty$.
In this limit we have
\begin{equation}\label{}
\!\!\!\!R=\lim_{z_0\to+\infty}\frac{\cosh(\pi(\omegA\!-\!\Ll))\cosh(\pi((2z_0\!-\!1)\omegA\!-\!\Ll))}{\cosh(\pi(\omegA\!+\!\Ll))
\cosh(\pi((2z_0\!-\!1)\omegA\!+\!\Ll))}=
 \frac{\cosh(\pi(\omegA\!-\!\Ll))\,e^{-2\pi\Ll}}{\cosh(\pi(\omegA\!+\!\Ll))}\,.
\end{equation}
Since one could doubt the legality of this limit because of the fact that $r_-=r_+$ is a singularity in our initial substitution (\ref{r-subst}) for $z$, and also for completeness of our investigation, we will consider this case separately.

In the extremal case  equation (\ref{R-wave}) becomes
\begin{equation}\label{}
\frac{d}{dr}\left[(r\!-\!r_+)^2\dRr\right]+\left[\frac{\omegA^2r^2}{(r\!-\!r_+)^2}-l(l\!+\!1)\right]\Rr=0  .
\end{equation}
The smaller number of singular points is a significant reason to require a separate investigation, so we cannot expect to obtain again hypergeometric equation.
In the extremal  case our new variable will be:
\begin{equation}\label{subst2}
    \;q:=\frac{r_+}{r\!-\!r_+}\quad\Rightarrow\quad\frac{d}{dr}=-\frac{q^2}{r_+}\frac{d}{dq}.
\end{equation}
So the equation for \ $\Qq:=\Rr$ takes the form:
$$q^2\ddQq+\left[\omegA^2\!-\!l(l\!+\!1)+2\omegA^2q+\omegA^2q^2\right]\Qq=0 . $$

Here the same substitution $\Ll:=\sqrt{\omegA^2\!-\!(l\!+\!1/2)^2}$ is appropriate for recognizing  the above equation
as Whittaker equation for $\Wz:=\Qq$, where \ $z:=2i\omegA q$:
\begin{equation}\label{WhittEQ}
\ddWz+\left[\frac{1/4-(i\Ll)^2}{z^2}+\frac{-i\omegA}z-\frac14\right]\Wz=0\,.
\end{equation}
The general solution can be represented in terms of the confluent hypergeometric functions (Kummer functions \cite{Abr}):
\begin{equation}\label{WhittEQ-sol}
\Wz=e^{-z/2}z^{\frac12+i\Ll}\left(C_1\,M(a,b,z)+C_2\,U(a,b,z)\right)\;,\;\;\textrm{where }
 \begin{array}{|l}
  a=\al,\\
  b=1\!+\!2i\Ll .
 \end{array}
\end{equation}

The asymptotic solution on the horizon ($r\to r_++0\Rightarrow q\to+\infty$) follows the asymptotic expansion of the Kummer functions \cite{Abr}:
$$\begin{array}{rl}\displaystyle
    \frac{M(a,b,z)|_{|z|\to+\infty}}{\Gamma(b)}\simeq&\!\!\displaystyle\frac{e^{\pm i\pi a}
z^{-a}}{\Gamma(b\!-\!a)}\left[\sum_{n=0}^{R-1}\frac{(a)_n(1\!+\!a\!-\!b)_n}{n!}(-z)^{-n}+O(|z|^{-R})\right]+\\
\vspace{-2ex}\\
    &\displaystyle\!\!+\frac{e^zz^{a-b}}{\Gamma(a)}\left[\sum_{n=0}^{S-1}\frac{(b\!-\!a)_n(1\!-\!a)_n}{n!}(-z)^{-n}+O(|z|^{-S})\right],
  \end{array}
$$
where \ $+i\pi a$ \ is for \ $-\pi/2\!<\!\arg z\!<\!3\pi/2$ and \ $-i\pi a$ \ is for \ $-3\pi/2\!<\!\arg z\!<\!-\pi/2$. We also have

$$U(a,b,z)|_{|z|\to+\infty} \simeq z^{-a}\left[\sum_{n=0}^{R-1}\frac{(a)_n(1\!+\!a\!-\!b)_n}{n!}(-z)^{-n}+O(|z|^{-R})\right]\,,
\;\;\left(-\frac{3\pi}2<\arg z<\frac{3\pi}2\right).$$

In our case the  zero order terms in the sums ($R=1, S=1$) are sufficient again. Taking
into account that $\arg z=\pi/2$ and also the expressions
(\ref{WhittEQ-sol}), one can obtain the solution on the horizon:

$$\begin{array}{r}
 \displaystyle Q(q\to+\infty) \simeq C_1\,\Gamma(1\!+\!2i\Ll)\!\left[\frac{ie^{-\pi(\Ll+\omegA)}e^{-i\omegA q}(2i\omegA q)^{-i\omegA}}
 {\Gamma(\bl[-])}+\frac{e^{i\omegA q}(2i\omegA q)^{i\omegA}}{\Gamma(\al)}\right]\,+\\
 \vspace{-2ex}\\
  +\,C_2\,e^{-i\omegA q}(2i\omegA q)^{-i\omegA}   .
\end{array}
$$

Here the suitable substitution that transforms the general solution in terms of 1D wave modes in $\mathbb R$ is:
\begin{equation}\label{xr2}
2qe^q:=e^{-x/r_0}\quad\Rightarrow\quad     x=\frac{-r_0\,r_+}{r\!-\!r_+}+r_0\ln\frac{r\!-\!r_+}{2r_+},
\end{equation}
which gives

\begin{equation}\label{A2}
\begin{array}{c}
  \Rr[r\to r_+] \simeq A_{out}e^{i\omega x}+A_{in}e^{-i\omega x},\\
  \vspace{-1ex}\\
  \displaystyle A_{out}:=C_1\frac{ie^{-\pi\Ll}e^{-\frac{\pi\omegA}2}\omegA^{-i\omegA}\,\Gamma(1\!+\!2i\Ll)}{\Gamma(\bl[-])}+C_2e^{\pi\omegA/2} \omegA^{-i\omegA}
\,,\;\
A_{in}:=C_1\,\frac{\Gamma(1\!+\!2i\Ll)e^{-\frac{\pi\omegA}2}\omegA^{i\omegA}}{\Gamma(\al)}.
\end{array}
\end{equation}

For the asymptotic solution at the radial infinity ($r\to+\infty\Rightarrow q\to0+\Rightarrow z\to0$)
we will use the relation between $M(a,b,z)$ and $U(a,b,z)$ \cite{Abr}:
$$U(a,b,z)=\frac\pi{\sin\pi b}\left[\frac{M(a,b,z)}
{\Gamma(1\!+\!a\!-\!b)\Gamma(b)}-z^{1-b}\frac{M((1\!+\!a\!-\!b,2\!-\!b,z)}{\Gamma(a)\Gamma(2\!-\!b)}\right]\,.$$

After applying the upper relation for $z=0$ where $M(a,b,0)=1$ we reach  the asymptotic solution at the spatial infinity.
$$Q(q\to 0)\simeq(2i\omegA q)^{1/2}\left\{\left[C_1+\frac{C_2\:i\pi}{\sinh(2\pi\Ll)\Gamma(\bl)\Gamma(1\!+\!2i\Ll)}\right](2i\omegA q)^{i\Ll}-\right.$$
$$\left.-\frac{C_2\,i\pi}{\sinh(2\pi\Ll)\Gamma(\al)\Gamma(1\!-\!2i\Ll)}(2i\omegA q)^{-i\Ll}\right\}.$$

The substitutions (\ref{xr2}) \ and \ $k_{\omega l}:=\Ll/r_0$ \ again give the asymptotic solution in the form of 1D wave in $\mathbb R$:

$$\Rr[r\to\infty] \simeq \sqrt{\frac{r_+}{r}} \left(B_{out}e^{ik_{\omega l}x}+B_{in}e^{-ik_{\omega l}x}\right) ,$$

where

\begin{equation}\label{B2}
\begin{array}{l}
 \displaystyle B_{out}=\frac{-C_2\,i\pi(2i\omegA )^{1/2}\,(i\omegA)^{-i\Ll}}{\sinh(2\pi\Ll)\Gamma(\al)\Gamma(1\!-\!2i\Ll)},\\
 \vspace{-2ex}\\
 \displaystyle B_{in}=\left[C_1+\frac{C_2\:i\pi}{\sinh(2\pi\Ll)\Gamma(\bl)\Gamma(1\!+\!2i\Ll)}\right]
 (2i\omegA )^{1/2}\,(i\omegA)^{i\Ll}.
\end{array}
\end{equation}

We should note that due to the time-reversal symmetry of wave equation it is permissible to work with the time-reversed formulation
of the radiation boundary condition. In the normal picture the radiation boundary condition
means missing of the in-going mode at the spatial infinity $B_{in}:=0$. The reflection coefficient in this case is $R:=|A_{in}|^2/|A_{out}|^2$.

In the time-reversed picture the radiation mode becomes an in-going mode so in this treatment $B_{out}:=0$. Respectively on the horizon the falling mode
becomes an in-going mode and the reflected mode becomes out-going mode. So in this case we should take $R:=|A_{out}|^2/|A_{in}|^2$.

It is possible to turn out that
$$\left.\frac{A_{in}}{A_{out}}\right|_{B_{in}:=0}\neq\left.\frac{A_{out}}{A_{in}}\right|_{B_{out}:=0}\;,\;\textrm{but always}
\;\;\left|\frac{A_{in}}{A_{out}}\right|^2_{B_{in}:=0}=\left|\frac{A_{out}}{A_{in}}\right|^2_{B_{out}:=0}\,.$$

In the work \cite{CFM} the time-reversed picture is chosen despite the fact that there is no difference in the complexity of further calculations between both approaches.
It is the same for our non-extremal black holes. But one could see from the expressions (\ref{A2}) and  (\ref{B2}) that the time-reversed formulation gives a
shorter way to $R$. All the results for $R$ were made by us using both approaches for checking the correctness of all previous calculations.
Following  the time-reversed picture $B_{out}=0 \Rightarrow\; C_2=0$ we find
\begin{equation}\label{}
 \begin{array}{r}
 \displaystyle R=\frac{|A_{out}|^2}{|A_{in}|^2}=\left|\frac{ie^{-\pi\Ll}e^{-\frac{\pi\omegA}2}\,\Gamma(1\!+\!2i\Ll)\omegA^{-i\omegA}}{\Gamma(\bl[-])}\right|^2
\left|\frac{\Gamma(\al)}{\Gamma(1\!+\!2i\Ll)e^{-\frac{\pi\omegA}2}\omegA^{i\omegA}}\right|^2=\\
\vspace{-2ex}\\
 \displaystyle =\frac{|\Gamma(\al)|^2}{|\Gamma(\bl[-])|^2}e^{-2\pi\Ll}=\frac{\cosh(\pi(\Ll\!-\!\omegA))}{\cosh(\pi(\Ll\!+\!\omegA))}e^{-2\pi\Ll}.
 \end{array}
\end{equation}

For high frequencies $\omegA\gg l\!+\!1/2 $ ( $\Rightarrow\Ll\approx\omegA$)  we have
\begin{equation}
R \approx \frac{e^{-2\pi\omegA}}{\cosh(2\pi\omegA)}\approx e^{-4\pi\omegA}\;\;\Rightarrow\;\;
N\approx e^{-4\pi\omegA}.
\end{equation}

Hence we find the Hawking temperature in the extremal case in
semiclassical approximation

\begin{eqnarray}
T_{H}= \frac{1}{4\pi r_{0}}.
\end{eqnarray}

Contra-intuitively the temperature of the extremal case is non-zero
and, as in the non-extremal case, is independent of the intrinsic
characteristics of the black hole and depends only on the background
parameter $r_{0}$.

\section{Discussion}
In the present paper we studied the Hawking radiation of
asymptotically non-flat dyonic black holes in 4D
Einstein-Maxwell-dilaton gravity  in semi-classical approximation.
It was shown that the problem can be solved exactly and we computed
exactly the semi-classical radiation spectrum of both non-extremal
and extremal black holes.

Our results show that the Hawking temperature, calculated in the
semi-classical approximation, is not compatible with the first law
(\ref{FLaw}). The  reason for this discrepancy  is  that the
spacetime is asymptotically non-flat. Other examples for discrepancy
between the Hawking temperature and the surface gravity  in
asymptotically non-flat spacetimes can be found in \cite{Park},
\cite{MHSG} and references therein. In principle, the relation
between the Hawking radiation and the first law in asymptotically
non-flat spacetimes is controversial and depends on the particular
case. In general, the first law in asymptotically non-flat spacetime
is not directly connected to the temperature of the particle flux at
infinity. For example, the black holes considered in the present
work can not emit massive particles because the mass term in the
wave equation (\ref{WE}) leads to the appearance of confining
potential (growing unboundedly to infinity) which prevents the
particles from escaping to infinity. In other words, the Hawking
temperature for massive particles, measured by an asymptotic
observer, is zero.

The asymptotical non-flatness leads to ambiguous thermodynamical
characteristics and thermodynamics as a whole. For example, the
surface gravity is given by the formula
\begin{eqnarray}
\xi^{\mu}\nabla_{\mu}\xi^{\nu}=\kappa \xi^{\nu}
\end{eqnarray}
on the horizon.  The above definition, however, gives the surface
gravity up to a constant because there is a freedom to rescale $\xi$
by a constant. In the asymptotically flat case this rescaling
freedom can be fixed by choosing a unit norm for the Killing field
at infinity. In linear dilaton spacetimes there is no natural way to
fix the rescaling freedom. The choice $\xi=\partial/\partial t$ made
in the papers devoted to the linear dilaton black holes is thus ad
hoc. The ambiguity in choosing the time vector field is present also
in the Hamiltonian formalism for linear dilaton spacetimes
\cite{CGLO} where the time vector field is again
$\xi=\partial/\partial t$.

One possible way to overcome the problem of the rescaling freedom in
our case is to replace the Killing field $\xi$ with the Kodama
vector field $K=\sqrt{\frac{r_{0}}{r}}\frac{\partial}{\partial t}$
which has a unit norm at infinity \cite{Kodama}. This investigation
is in progress and the results will be presented elsewhere.


\vspace{1.5ex}
\begin{flushleft}
\large\bf Acknowledgments
\end{flushleft}

The partial financial supports from the  Bulgarian National Science
Fund under Grant DMU-03/6, and by Sofia University Research Fund
under Grant 148/2012 are gratefully acknowledged.

\end{document}